\documentstyle[12pt]{article}
\begin{document}
\begin{titlepage}
\title{A True Equation to Couple Classical and Quantum Dynamics}
\author{Lajos Di\'osi\thanks{E-mail: diosi@rmki.kfki.hu}\\
KFKI Research Institute for Particle and Nuclear Physics\\
H-1525 Budapest 114, POB 49, Hungary\\\\
{\it e-print archives ref.: quant-ph/9510028\hfill}}
\date{October 27, 1995}
\maketitle
\begin{abstract}
Starting from the Schr\"odinger-equation of a composite system, we derive
unified dynamics of a classical harmonic system coupled to an arbitrary
quantized system. The classical subsystem is described by random phase-space
coordinates entangled with the quantized variables of the complementary
subsystem. Our semiclassical equation is {\it true} in a sense that its
predictions are identical to those of the fully quantized composite dynamics.
This exact method applies to a broad class of theories, including e.g. the
relativistic quantum-electrodynamics and the electron-fonon dynamics.
\end{abstract}
\end{titlepage}

{\it Introduction\/.}
A consistent model for dynamical coupling between classical and quantum
systems would be attractive for various theoretical and practical reasons.
We shall deliberately call such model {\it semiclassical}. It would,
for instance, explain the "collapse" of wave function during quantum
measurement
as a result of dynamical interaction between the classical
measuring apparatus and the measured quantum system \cite{SheSud}.
Furthermore, particle physics would be understood on the classical background
of space-time \cite{Ros}. A remarkable theoretical proposal
of a canonical semiclassical dynamics \cite{Ale} was motivated by more
practical aims, related to physics of molecules.
Unfortunately,
in Ref.~\cite{SheSud}, some variables of the classical subsystem acquire
quantum uncertainties. The mean-field model \cite{Ros} is,
on the contrary, unable
to explain the obvious statistical uncertainties arising in the
classical subsystem's dynamics due to the backreaction of the quantum
subsystem.
The canonical formalism \cite{Ale} may not preserve the
positivity of state distribution function, cf. Ref.~\cite{BouTra}.
As yet, all speculations have failed to obtain a consistent semiclassical
model.

For the time being, we are not able to model the backreaction of a
quantum system onto a generic classical one. Here, nevertheless,
harmonic oscillator is a favorable exception: its quantum and classical
dynamics map onto each other exactly \cite{PopPer}.
Classical phase-space coordinates appear as labels of
coherent quantum states \cite{Gla} which form an overcomplet basis
for the oscillator's Hilbert-space.
One inclines therefore to guess that classical description could be used for
a system of quantum harmonic oscillators ({\it harmonic system\/}, in short)
even if it interacts dynamically with another quantum system, provided the
interaction is linear or at most quadratic in the harmonic variables.
In our Letter we show that, using coherent state representation,
this is indeed
the case.

We derive semiclassical equation for the wave function
of a composite system containing a harmonic subsystem.
This subsystem is described in classical {\it stochastic\/} terms while it
interacts with the {\it quantized} dynamics of the complementary subsystem.
Our semiclassical equation provides,
by quantum {\it and\/} stochastic averages, the true
quantum expectation values for the fully quantized composite system operators.

{\it The model.\/}
We consider a quantum system consisting of two interacting subsystems.
One of them is assumed to be harmonic. Its interaction with the other
subsystem should be linear in the harmonic variables.
This class includes
many systems of primary physics interest like, e.g., quantum-electrodynamics
(hence quantum-optics as well), the electron-fonon interaction,
the spin-boson system \cite{Leg},
or the so-called Caldeira-Leggett model \cite{Cal}.
We shall use the terminology of quantum-electrodynamics
or quantum-optics whenever concrete terms help explanation.
Accordingly, we consider the following form of Hamiltonian:
\begin{equation}
H=H_0+\sum_n \omega_n a_n^\dagger a_n
	+\sum_n\left(J_n a_n^\dagger+J_n^\dagger a_n \right)
\end{equation}
where the Planck's constant $\hbar$ has been set to $1$.
The $a_n,a_n^\dagger$ are absorption and creation operators for the
harmonic system (which will be called {\it field\/}).
The $\omega_n$'s are the field mode frequencies.
$H_0$ stands
for the Hamiltonian of the charged particles (which we call {\it atomic
system\/},
for simplicity's). The atomic operators $J_n$ coupled to the field modes
will be called {\it currents\/}. In interaction picture, we get the following
Schr\"odinger-equation for the state vector $\Psi$ of our composite
quantum system:
\begin{equation}
\dot\Psi_t
=-i\sum_n\left(j_n(t) a_n^\dagger+j_n^\dagger(t) a_n \right)\Psi_t
\end{equation}
where $j_n(t)=\exp(i\omega_n t)J(t)$ is the "rotated" version of the
interaction picture current $J_n(t)$ for each field mode.

Instead of the abstract state vector $\Psi$,
we introduce a wave function in {\it coherent state\/} representation
for the harmonic subsystem, i.e. for the field:
\begin{equation}
\psi(\alpha,\alpha^\ast)=\langle \alpha,\alpha^\ast \vert \Psi \rangle,
\end{equation}
where
$\vert \alpha,\alpha^\ast \rangle$ is equal to the
direct product of the coherent states
$\vert \alpha_n,\alpha_n^\ast\rangle$ of all field oscillators
$(n=1,2,\dots)$, respectively.
The scalar product on the RHS is to be taken on the factor Hilbert-space
of the field. (The author has failed to find a more suitable compact
notation.) Accordingly, $\psi(\alpha,\alpha^\ast)$
is wave function of the field
coordinates $\alpha,\alpha^\ast$ while, on the other hand, it is abstract
state vector in the atomic Hilbert-space.

{\it Conditional quantum state.\/}
We are going to give a special
interpretation to the wave function (3).
Let us consider the norm function which is known as the $Q$-function
in quantum optics \cite{Gar}:
\begin{equation}
w(\alpha,\alpha^\ast)=\Vert\psi(\alpha,\alpha^\ast)\Vert^2.
\end{equation}
Formally, it is a probability distribution
over the classical phase-space of the field. From now on,
let us choose this way to interpret it! Consequently, the field will
be considered classical and stochastic, described by random phase-space
coordinates $\alpha,\alpha^\ast$.

Given an
arbitrary function $F(\alpha,\alpha^\ast)$, its stochastic mean will
reproduce the exact expectation value of the
corresponding antinormally ordered operator
$:\!\!F(a,a^\dagger)\!\!:_{anti}$ as calculated in the original
quantum state, i.e.:
\begin{equation}
\Psi^\dagger\!:\!\!F(a,a^\dagger)\!\!:_{anti}\!\Psi=
\int F(\alpha,\alpha^\ast)w(\alpha,\alpha^\ast)\prod_n {d^2\alpha_n\over\pi}.
\end{equation}
This relation follows from standard literature on coherent states
\cite{Gar}.
It is straightforward to inspect that a more general relation holds true
in cases where $F(\alpha,\alpha^\ast)$ is not necessarily an
$(\alpha,\alpha^\ast)$-dependent {\it number\/} but an arbitrary
atomic {\it operator\/}.
Then we can write:
\begin{equation}
\Psi^\dagger\!:\!\!F(a,a^\dagger)\!\!:_{anti}\!\Psi=
\int\psi^\dagger(\alpha,\alpha^\ast)F(\alpha,\alpha^\ast)
\psi(\alpha,\alpha^\ast)
\prod_n {d^2\alpha_n\over\pi}.
\end{equation}

Invoking the Eq.~(4) and its probabilistic interpretation suggested above,
it is reasonable to introduce the notion of {\it conditional quantum
expectation value\/} of a given atomic operator $F$:
\begin{equation}
\bigl\langle F \bigr\rangle_{\alpha,\alpha^\ast}
\equiv{\psi^\dagger(\alpha,\alpha^\ast) F \psi(\alpha,\alpha^\ast)
	\over\Vert\psi(\alpha,\alpha^\ast)\Vert^2},
\end{equation}
which, in fact, is the quantum expectation value of $F$ at fixed
state $(\alpha,\alpha^\ast)$ of the classical field. Lastly but most
importantly, we shall interpret the wave function
$\psi(\alpha,\alpha^\ast)$ as {\it conditional quantum state\/}
of the atomic subsystem at fixed classical state of the field. That this way
of interpretation is consistent we see by applying Eqs.~(4) and (7) on
the RHS of Eq.~(6):
\begin{equation}
\Psi^\dagger\!:\!\!F(a,a^\dagger)\!\!:_{anti}\!\Psi=
\int\bigl\langle F(\alpha,\alpha^\ast)\bigr\rangle_{\alpha,\alpha^\ast}
w(\alpha,\alpha^\ast)\prod_n {d^2\alpha_n\over\pi}.
\end{equation}

In summary, the atom's state is described by state vector while
the field's state is characterized by random phase-space coordinates
$(\alpha,\alpha^\ast)$. The state of the composite system is given
by the unnormalized conditional atomic state
$\psi(\alpha,\alpha^\ast)$ whose norm yields the phase-space distribution
of the field. The original quantum expectation values of
entangled atomic and normal ordered field operators can exactly be obtained
by stochastic averaging of the corresponding conditional atomic quantum
expectation values over all classical fields.

{\it Equation of motion.\/}
In coherent state representation the following operator correspondences
should be used (cf. Ref.~\cite{Gar}):
\begin{equation}
a_n^\dagger\Psi\leftrightarrow \alpha_n^\ast\psi(\alpha,\alpha^\ast),~~~~~
a_n\Psi\leftrightarrow
\left({\alpha_n\over2}+{\partial\over\partial \alpha_n^\ast}\right)
\psi(\alpha,\alpha^\ast).
\end{equation}
Hence the Schr\"odinger-equation (2) leads to the following
equation of motion for the wave function (3):
\begin{equation}
\dot\psi_t(\alpha,\alpha^\ast)=
-i\left(\alpha^\ast j(t) + {\alpha\over2} j^\dagger(t)\right)
\psi_t(\alpha,\alpha^\ast)
-i j^\dagger(t){\partial\psi_t(\alpha,\alpha^\ast)\over\partial \alpha^\ast},
\end{equation}
where we have suppressed the notation of indices $n$ as well as of summation
over them.

{}From the standard theory of coherent states it follows that the conditional
state vector (3) can always be written in the following form:
\begin{equation}
\psi(\alpha,\alpha^\ast)
=e^{-\vert\alpha\vert^2/2}\varphi(\alpha^\ast).
\end{equation}
Since the so-called Bargmann-state $\varphi(\alpha^\ast)$ \cite{Bar} does not
depend on $\alpha$ the equation of motion (10) becomes simpler in the
Bargmann-representation:
\begin{equation}
\dot\varphi_t(\alpha^\ast)=
	-i\alpha^\ast j(t)\varphi_t(\alpha^\ast)
	-i j^\dagger(t)
	{\partial\varphi_t(\alpha^\ast)\over\partial \alpha^\ast}.
\end{equation}

{\it Discussion.\/}
To interpret the interaction between the
the quantized atomic system and
the classical field, respectively,
we rewrite the equation of motion (10) [or (12), equivalently] for the
{\it conditional density operator\/} defined by
$\rho(\alpha,\alpha^\ast)\equiv
 \psi(\alpha,\alpha^\ast)\psi^\dagger(\alpha,\alpha^\ast)$.
Using the Eq.~(11), too, one obtains:
\begin{equation}
\dot\rho_t=
-i\left[\alpha^\ast j(t) + \alpha j^\dagger(t),\rho_t\right]
-ij^\dagger(t){\partial\rho_t\over\partial\alpha^\ast}
+i{\partial\rho_t\over\partial\alpha}j(t).
\end{equation}
The first term on the RHS is a Hamiltonian contribution to the quantum state
dynamics as if the classical field were an external parameter.
The derivative terms differ by operator orderings from the terms proposed
in Ref.~\cite{Ale}. The difference leads to consistency.
These derivative terms represent a specific kind of backreaction
of quantum fluctuations, controlling the stochastic spread of the
classical system's variables. To see that, let us trace both sides  of
the equation over the atomic Hilbert-space. We obtain
the following {\it drift\/} equation for the phase-space density (4):
\begin{equation}
\dot w_t(\alpha,\alpha^\ast)
=-i{\partial\over\partial\alpha^\ast}
	\left(\bigl\langle j^\dagger(t) \bigr\rangle_{\alpha,\alpha^\ast}
		w_t(\alpha,\alpha^\ast)\right) + c.c.~~~~.
\end{equation}
Hence, the phase-space coordinates of each member in the random ensemble
obey to {\it deterministic\/} classical Hamiltonian equations of motion
\begin{equation}
\dot\alpha =-i\bigl\langle j(t) \bigr\rangle_{\alpha,\alpha^\ast}~~~,~~~~
\dot\alpha^\ast
=i\bigl\langle j^\dagger(t) \bigr\rangle_{\alpha,\alpha^\ast}~~~,
\end{equation}
where quantized atomic currents appear via their conditional quantum
mean values. These equations coincide formally with the mean-field
equations (cf.,e.g., in Refs.~\cite{Ros,BouTra}). In mean-field theory
the classical phase-space variables take sharp values. It does not
include the backreaction of the  fluctuations in the quantum subsystem.
In our theory, however, the influence of those fluctuations is reflected
by the stochasticity of classical phase-space coordinates.

{\it Summary, outlook.\/}
We derived dynamic coupling between a classical harmonic
system and a generic quantum system.
Our semiclassical equation yields, via quantum and stochastic averages,
all quantum expectation values of the full quantized dynamics of the
composite system. The scope of this semiclassical theory is not
restricted to the non-relativistic quantum mechanics.
It applies to a broad class of theories, including first of all
the relativistic quantum-electrodynamics, the electron-fonon interaction
in solid states, the spin-boson problem etc.

Admittedly, our claim that exact quantum results may be computed from
stochastic averages would provoque concerns. Especially,
the existence of a stochastic model for the radiation field
interacting with quantized currents raises conceptual issues
such as, e.g., a possible conflict with Bell's theorem which forbids the
existence of local stochastic models. Here we do not intend to discusse
the corresponding loopholes.
Some measurement theoretical implications have already been discussed
elsewhere \cite{Dio}.

We recall the fundamental issues
mentioned in the introduction as main motivations. The collapse of wave
function can certainly be described if we apply our equations to a
suitably chosen model of measurement situation. Most interestingly, a natural
collapse dynamics appears whenever we let our quantum system to interact
with the classical electromagnetic and/or gravitational fields, described
of course in the mathematical framework of the present Letter.
Finally, we emphasize the practical value of the possibility that a
great deal of interacting quantum systems becomes exactly treatable in
semiclassical framework which is in many respect simpler than the original
quantum one.

\vskip .5truecm
{\it Appendix: Principles of numeric simulation.\/}
We outline the principles of an efficient algorithm to simulate the solutions
of the semiclassical equations of motion.
Let us assume that we know the (Bargmann-) conditional
state $\varphi_t(\alpha^\ast)$ at some initial time $t$.
Then the initial phase-space distribution of the field is
\begin{equation}
w_t(\alpha,\alpha^\ast)
=e^{-\vert\alpha\vert^2}\Vert\varphi_t(\alpha^\ast)\Vert^2.
\end{equation}
For the first numeric step, we generate a long random sequence
of complex phase-space coordinates
$\alpha(1),\alpha(2),\dots,\alpha(N)$ according to the distribution (16).
Meanwhile, we keep the steps  $\vert\alpha_n(k+1)-\alpha_n(k)\vert$
small for all $n$ and $k$. (Especially in case of many field modes,
the Monte-Carlo "importance sampling" \cite{Met} seems useful to generate
such random sequences.)
We store the sequence $\{\alpha(k);k=1,N\}$. Then we calculate
$\varphi(k)\equiv\varphi_t(\alpha^\ast(k))$ for each $k$
and store the sequence $\{\varphi(k);k=1,N\}$, too.
These two sequences will represent the state of our system numerically.
Their time evolution will be deterministic. Let us calculate their values
for time $t+\epsilon$ where the time increment $\epsilon$ is small.
{}From the equation of motion (15) we read out the updated phase-space
coordinates:
\begin{equation}
\alpha_n(k)\rightarrow
\alpha_n(k)-i\epsilon {\varphi^\dagger(k)j_n(t)\varphi(k)\over
		     \Vert\varphi(k)\Vert^2},~~~~(k=1,2,\dots,N;n=1,2,\dots).
\end{equation}
The numeric approximation of the equation of motion (12) leads to the
following update of the conditional state vectors:
\begin{equation}
\varphi(k)
\rightarrow\varphi(k)-i\epsilon\sum_n\alpha_n^\ast(k)j_n(t)\varphi(k)
		     -i\epsilon \sum_n j_n^\dagger(t)
{\varphi(k\pm1)-\varphi(k)\over\alpha_n^\ast(k\pm1)-\alpha_n^\ast(k)},
\end{equation}
for $n=1,2,\dots$. The upper sign is to be taken for $k=1,2,\dots,N-1$;
we take the lower one for $k=N$.
Repeated applications of the basic update cycle (17,18) yield the
state of the system for any later time. The quantum expectation
value of any (antinormally ordered) composite operator can be
calculated numerically.
To calculate quantum expectation values, we approximate the RHS
of the analytic relation  (7) numerically as follows:
\begin{equation}
\Psi^\dagger\!:\!\!F(a,a^\dagger)\!\!:_{anti}\!\Psi\approx
{1\over N}\sum_{k=1}^N
{\varphi^\dagger(k)F(\alpha(k),\alpha^\ast(k))\varphi(k)
 \over\Vert\varphi(k)\Vert^2}
\end{equation}
i.e., it is approximately equal to the ensemble average of the
corresponding conditional quantum expectation values. The relation
becomes identity in the limit $N\rightarrow\infty$.

The numeric derivation
\begin{equation}
{\partial \varphi_t(\alpha^\ast(k))\over \partial \alpha_n^\ast}
\approx{\varphi_t(\alpha^\ast(k\pm1))-\varphi_t(\alpha^\ast(k))
\over\alpha_n(k\pm1)-\alpha_n(k)}
\end{equation}
is, as usual, an awkward step of the algorithm.
We have initially chosen small enough increments
$\vert\alpha_n(k+1)-\alpha_n(k)\vert$ to assure a good quality
of numeric derivation. However, the "chain" $\{\alpha(k);k=1,2,\dots,N\}$
may, after many updates, becomes distorted is such a way that its
increments are not uniform and small anymore. Then, a special
formatting procedure must be applied to both sequences
$\{\alpha(k)\}$ and $\{\varphi(k)\}$ to restore uniform small increments
before the updating steps continue.

\bigskip
The author is indebted to J. Butterfield for his illuminating remark on
conditional quantum states.
This work was supported by the grants OTKA No. 1822/1991 and T016047.


\begin{thebibliography}{99}
\bibitem{SheSud} T.N.Sherry and E.C.G.Sudarshan,
	Phys. Rev. {\bf D18}, 4580 (1978).
\bibitem{Ros} L.Rosenfeld, Nucl.Phys. {\bf 40}, 353 (1963).
\bibitem{Ale} I.V.Aleksandrov, Z.Naturf. {\bf 36a}, 902 (1981).
\bibitem{BouTra} W.Boucher and J.Traschen, Phys. Rev. D {\bf 37}, 3522 (1988).
\bibitem{PopPer} B.C.Popov and A.M.Perelomov, ZHETF {\bf 56}, 1375 (1969);
	{\bf 57}, 1684 (1969);
	see also in A.I.Baz, Ya.B.Zeldovich, and A.M.Perelomov:
	{\it Scattering, Reaction, and Decay in Non-relativistic
	Quantum Mechanics} (Nauka, Moscow, 1971, in Russian).
\bibitem{Gla} R.J.Glauber, Phys.Rev. 131, 2766 (1963).
\bibitem{Leg} A.J.Leggett {\it et al.\/}, Rev.Mod.Phys. {\bf 59}, 1 (1987).
\bibitem{Cal} A.Caldeira and A.J.Leggett, Physica {\bf A 121}, 587 (1983).
\bibitem{Gar} C.W.Gardiner, {\it Quantum Noise\/} (Springer, Berlin, 1991).
\bibitem{Bar} V.Bargmann, Commun.Pure Appl.Math. {\bf 14}, 187 (1961).
\bibitem{Dio} L.Di\'osi, Quant.Semiclass.Opt. (to appear);
	e-print archives ref.: quant-ph/9507010.
\bibitem{Met} N.Metropolis {\it et al.\/}, J.Chem.Phys. {\bf 21}, 1087 (1953);
K.Binder, in {\it Monte Carlo Methods in Statistical Physics\/},
	ed. K.Binder (Springer, Berlin, 1979).
\end{thebibliography}
\end{document}